\documentclass[12pt]{iopart}
\usepackage{iopams}  
\newcommand{\dket}[1]{| \, #1 \rangle\!\rangle}
\newcommand{\dbra}[1]{\langle\!\langle #1 \, |}
\begin{document}

\title[Characterization of tomographically faithful states in terms of
their Wigner function]{Characterization of tomographically faithful
states in terms of their Wigner function}

\author{G M D'Ariano 
\footnote[3]{To whom
correspondence should be addressed.} and M F Sacchi}

\address{QUIT, INFM and Dipartimento di Fisica ``A. Volta'',
  Universit\`a di Pavia, via A. Bassi 6, I-27100 Pavia, Italy}

\begin{abstract}
A bipartite quantum state is {\em tomographically faithful} when it
can be used as an input of a quantum operation on one of the two
quantum systems, such that the joint output state carries a complete
information about the operation itself.  Tomographically faithful
states are a necessary ingredient for tomography of quantum operations
and for complete quantum calibration of measuring apparatuses. In this
paper we provide a complete classification of such states for
continuous variables in terms of the Wigner function of the state.
For two-mode Gaussian states faithfulness simply resorts to
correlation between the modes.
\end{abstract}


\section{Introduction}
Quantum operations describe any kind of physical process affecting
quantum states, including unitary evolutions of closed systems and non
unitary transformations of open quantum systems, such as systems
interacting with a reservoir, or subjected to noise or measurements of
any kind.  The problem of determining experimentally the quantum
operation occurs in different scenarios, typically for quantum
calibration of controlled transformations \cite{tomo_channel} and of
measuring apparatuses \cite{calib}.

\par In a {\em naive} process tomography one varies the input state
over a suitably complete set in order to recover enough information
about the quantum operation. The tensor structure of the bipartite
quantum system, however, allows to use a {\em single} fixed bipartite
state that scans the complete set of single-system states in a quantum
parallel fashion \cite{tomo_channel}.  The bipartite states that can
be used in this way in order to carry a complete information of the
process are called {\em tomographically faithful} \cite{prl}.

\par The problem of evaluating the faithfulness of a state can be
expressed in terms of an invertibility condition of a map associated
to the state. In many situations it is not simple to check such a
condition. In this paper we address the {\em continuous variables}
case (i.e. quantum harmonic oscillators), and solve the problem of the
complete classification of faithfulness in terms of the Wigner
function of the state. The use of Wigner functions has proved very
useful as a generalized phase-space technique to express density
operators in terms of $c$-number functions, thus leading to a
considerable simplification of the evaluation of quantum dynamics and
of expectation values \cite{phs,gard}.

\par In this paper we present a general result that provides a
necessary and sufficient condition for the faithfulness in terms of
Wigner function. Such a condition, as we will show, makes use of
highly irregular functions, as the customary $P$-functions in quantum
optics. We then specialize our results to the case of Gaussian states
of two modes of the electromagnetic field. The class of Gaussian
states constitutes a fortunate framework both for theoreticians and
experimentalists, since, on one side all calculations can be done
analytically, whereas, on the other side, these states are easily
generated in a lab, using lasers, linear optics, and parametric
amplifiers. We will show that for Gaussian states the condition of
faithfulness is just the existence of correlations between the two
modes.

\par The paper is organized as follows. In Sec. II we briefly recover
the general result about the faithfulness of a quantum state, and
recall the problem of inversion of a special operator associated to
the state. We then restate the problem in terms of Wigner function of
the state, and write a necessary and sufficient condition.  The
section presents some examples of faithful (both entangled and
separable) and unfaithful states. In Sec. III we simplify the result
of Sec. II for the case of two-mode Gaussian states. The condition of
faithfulness then simply restates as the existence of correlations
between the two modes.  We conclude the paper in Sec. IV, with a
summary of results and some remarks about the statistical errors that
affect the reconstruction of a quantum operation, and the connection
with quantum images.

\section{Faithfulness in terms of Wigner function}
In mathematical terms, a quantum operation $\cal E$ is described by a
completely positive map \cite{Kraus}. This can be written in the Kraus
form
\begin{equation}
{\cal E}(\rho )= \sum _n K_n \rho K_n ^\dag  ,
\end{equation}
where $K_n$ are operators on the Hilbert space $\cal H$ of the quantum
system. For simplicity we will consider quantum operations with the
same input and output space $\cal H$, and that are
trace-preserving---the so-called {\em channels}---corresponding to the
completeness relation $\sum _n K^\dag _n K_n =I$. The concept of {\em
tomographically faithful} state \cite{prl} relies on using a bipartite
state $R $ on ${\cal H}\otimes {\cal H}$, such that the output state
\begin{equation}
R_{\cal E}= ({\cal E}\otimes I) R 
\end{equation}
is in one-to-one correspondence to the quantum operation $\cal E$. In
\cite{prl} it was proved that a state $R$ is faithful iff the
following operator on ${\cal H}\otimes {\cal H}$
\begin{equation}
\check R= (ER)^{\tau _2}E=(R^{\tau _2}E)^{\tau _1}\label{rcek}
\end{equation}
is invertible. In Eq. (\ref{rcek}) $E=\sum_{i,j}|ij \rangle \langle
ji|$ denotes the swap operator, and $O^{\tau _l}$ denotes the partial
transposition of the operator $O$ on the $l$th Hilbert space, $l=1,2$.
\par Using the notation of Ref. \cite{pla} for bipartite vectors
\begin{equation}
\dket{A}\equiv \sum_{n,m} \langle n|A|m \rangle \,|n \rangle \otimes
|m \rangle ,\label{nota}
\end{equation}
one can generally write a bipartite state in the form
\begin{equation}
R =\sum_{i,j} \dket{A_i}\dbra {B_j} .\label{xi}
\end{equation}
From the identity
\begin{equation}
A\otimes B \dket{C}= \dket{ACB^\tau } \label{abc}
\end{equation}
it follows that 
\begin{equation}
\check R=\sum _{i,j} A^\tau _i \otimes B^\dag _j ,\label{tra}
\end{equation}
where the transpose $\tau $ is defined on the basis chosen in the
decomposition of Eq. (\ref{nota}).  Similarly, for a state $R$ written
as
\begin{equation}
R=\sum _{i,j} A_i \otimes B_j ,\label{8}
\end{equation}
using Eq. (\ref{rcek}), one easily writes $\check R$ as follows
\begin{equation}
\check R= \sum_{i,j}\dket{B_j}\dbra {A_i ^*}
 ,\label{scek} 
\end{equation}
where $O ^*$ denotes the complex conjugation on the fixed
basis. Notice that the evaluation the operator $\check R$ does not
need the diagonalization of $R $.  Moreover, all the previous sums can
be suitably replaced with integrals.

\par A state $R$ is faithful iff the associated operator $\check R$ is
invertible. We are interested in finding the conditions of
faithfulness in terms of the Wigner function of the state. For
simplicity, we consider bipartite states that correspond to two-mode
states. However, our results are easily generalized to the case of
two-party multimode states.  \par We recall the Cahill-Glauber
formulas \cite{gla} between single-mode density matrix $\rho $ and
Wigner function $W(\alpha ,\alpha ^*)$
\begin{eqnarray}
&&W(\alpha ,\alpha ^* ) =\frac {2}{ \pi }\hbox{Tr }[\rho D(2\alpha)
(-1)^{a^\dag a}] ,\\ &&\rho =2 \int _{\mathbb C} d^2 \alpha \,W(\alpha
,\alpha ^* )D(2 \alpha ) (-1)^{a^\dag a},\label{Wigner}
\end{eqnarray}
where $\alpha ^* $ denotes the complex conjugate of $\alpha $, $d^2
\alpha \equiv d\,\hbox{Re}(\alpha )\,d\,\hbox{Im}(\alpha )$, and
$D(\alpha )=e^{\alpha a^\dag -\alpha ^* a }$ represents the
displacement operator for the mode $a$, with $[a, a^\dag ]=1$.

\par For a two-mode bipartite state $R$ we
write the Wigner function as a function of two complex variables
$\alpha$ and $\beta$ by direct generalization of Eq. (\ref{Wigner}) as
follows
\begin{equation}
R = 4 \int _{\mathbb C} d^2 \alpha \int _{\mathbb C} d^2 \beta
\,W(\alpha ,\alpha ^* , \beta ,\beta ^* ) D(2 \alpha )(-1)^{a^\dag a}
\otimes D(2 \beta )(-1)^{b^\dag b},
\label{rwig}
\end{equation}
where $a$ is a shorthand notation for the annihilation operator
otherwise denoted $a\otimes I$, as well as $b$ for $I\otimes b$.  In
the following we will write the Wigner function shortly as
$W(\alpha,\beta )\equiv W(\alpha ,\alpha ^* , \beta ,\beta ^* ) $,
omitting the dependence on the complex conjugated variables.
According to Eqs. (\ref{8}) and (\ref{scek}), the condition of
faithfulness of the state in (\ref{rwig}) corresponds to the condition
of invertibility of the operator
\begin{eqnarray}\!\!\!\!\!\!\!\!\!\!\!\!\!\!\!
\check R  &= &
4 \int _{\mathbb C} d^2 \alpha \int _{\mathbb C} d^2 \beta \,W(\alpha
, \beta )\, 
\dket{D(2 \beta  )(-1)^{a^\dag a} }\dbra{D(2 \alpha ^* )(-1)^{a^\dag
  a}} \\ \!\!\!\!\!\!\!\!\!\!\!\!\!\!\! &= &
4 (I\otimes (-1)^{b^\dag b})\int _{\mathbb C} 
d^2 \alpha \int _{\mathbb C} d^2 \beta \,W(\alpha
, \beta )\, 
\dket{D(2 \beta  )}\dbra{D(2 \alpha ^* )} (I\otimes (-1)^{b^\dag b})
\;,\nonumber 
\end{eqnarray}
where in the second line we used identity (\ref{abc}).  Since the set
$\{ \dket{D(\alpha )} \}$ is an orthonormal basis (in the Dirac sense) for
${\cal H}\otimes {\cal H}$, namely $\int _{\mathbb C}\frac{d^2 \alpha
}{\pi }\dket {D(\alpha )}\dbra {D(\alpha )}=I\otimes I$, the condition
of faithfulness in terms of Wigner function is then the following:

\paragraph{\bf Necessary and sufficient condition for faithfulness:} 
a bipartite state with Wigner function $W(\alpha ,\beta )$ is faithful
iff one can find a function $f(\beta , \gamma )$ such that
\begin{equation}
\int _{\mathbb C}d^2 \beta \,W(\alpha ,\beta ) \,f(\beta ,\gamma )= 
\delta ^{(2)}(\alpha -\gamma ) \;,\label{condw}
\end{equation}
where $\delta ^{(2)}(\sigma )=\int _{\mathbb C}\frac {d^2 \lambda
}{\pi^2 }\,e^{\lambda \sigma ^* -\lambda ^* \sigma }$ denotes the
Dirac delta over the complex plane.  \par Equation (\ref{condw})
should be read in distributional sense. When such a condition is
satisfied one has
\begin{equation}\!\!\!\!\!\!\!\!\!\!\!\!\!\!\!\!\!\!\!
\check R ^{-1}=\frac {4}{\pi} 
 (I\otimes (-1)^{b^\dag b})\int _{\mathbb C} 
d^2 \delta  \int _{\mathbb C} d^2 \gamma  \,f(\delta ,\gamma )\,
\dket{D(2 \gamma  ^*) }\dbra{D(2 \delta  )} (I\otimes (-1)^{b^\dag  b}) \;.
\end{equation}
\subsection{Example 1: twin-beam state}
Consider the twin-beam state that can be easily generated by
nondegenerate optical parametric amplifiers
\begin{equation}
R=(1- \lambda ^2)\dket{\lambda ^{a^\dag a}}\dbra{\lambda ^{a^\dag
    a}} \;,\qquad 0\leq \lambda < 1 \;,
\end{equation}
where the parameter $\lambda $ is simply related to the total number
of photons $\bar n= 2 \lambda ^2/(1-\lambda ^2)$.  The corresponding
Wigner function is given by
\begin{equation}
W_R(\alpha ,\beta )=\frac {4 (1-\lambda ^2)}{\pi^2}\hbox{Tr} [\lambda
^{a^\dag a}D(2 \alpha ) \lambda ^{a^\dag a} D(-2 \beta ^*) ]\;
.\label{wr}
\end{equation}
By normal ordering Eq. (\ref{wr}) and a lengthy calculation, one
obtains
\begin{equation}
W_R(\alpha ,\beta )=\frac {4}{\pi^2}\exp \left[-\frac {2(1 +
\lambda^2)}{1-\lambda ^2}(|\alpha |^2 +|\beta |^2) +\frac{4 \lambda
}{1-\lambda ^2}(\alpha \beta + \alpha ^* \beta ^*) \right] \;.
\end{equation}
Using the solution (\ref{solu}) of identity (\ref{condw}) derived in
the appendix, the function $f(\beta , \gamma )$ can formally be
written as
\begin{eqnarray}
\!\!\!\!\!\!\!\!\!\!\!\!\!\!\!\!\!\!\!\!\!\!\!\!\!\!  f(\beta ,\gamma
)= \frac{4 \lambda ^2 }{\pi (1-\lambda ^2)^2}\,e^{\frac{1+3 \lambda
^2} {1-\lambda ^2}|\beta |^2}\,e^{2\frac{1-8 \lambda ^2 +\lambda
^4}{(1- \lambda ^2)^2} |\gamma |^2} \int _{\mathbb C}d^2 \xi \,
e^{|\xi |^2 }\, e^{\xi (\frac {4 \lambda }{1- \lambda ^2}\gamma -\beta
^*) - \xi ^*(\frac {4 \lambda }{1- \lambda ^2}\gamma ^*-\beta
)}\;,\label{d1}
\end{eqnarray}
which should be treated as a distribution, in the sense that the
integral in $\xi $ has to be performed after the integration on $\beta
$ of Eq. (\ref{condw}).  \par Notice that in this simple example, the
faithfulness is more easily checked using Eq. (\ref{tra}) and writing
immediately
\begin{eqnarray}
\check R ^{-1}= \frac {1}{1 - \lambda ^2 } \left (\frac {1}{\lambda
}\right ) ^{a^\dag a} \otimes \left (\frac {1}{\lambda }\right
)^{b^\dag b} \;.
\end{eqnarray}

\subsection{Example 2: classically correlated coherent states}
A mixture of correlated coherent states can be easily generated by
splitting thermal radiation in a $50/50$ beam splitter. For such a
kind of states we can write
\begin{equation}
R=\int _{\mathbb C}\frac {d^2 \gamma }{\pi \sigma ^2}\,
e^{-\frac{|\gamma |^2}{\sigma ^2}} |\gamma \rangle \langle \gamma
|^{\otimes 2},\label{corcoh}
\end{equation}
where the variance $\sigma $ is related to the total number of photons
by $\bar n = \sigma /2$.  The corresponding Wigner function is given
by
\begin{equation}\!\!\!\!\!\!\!\!\!\!\!\!\!\!\!\!\!\!\!\!\!\!\!\!\!\!\!\!\!\!
W_R(\alpha ,\beta )=\frac {4}{\pi^2(1+2 \sigma ^2)} \exp \left[-\frac
{2}{1+ 2 \sigma ^2}(|\alpha |^2 +|\beta |^2) +\frac{4 \sigma ^2 }{1 +2
\sigma ^2}(\alpha \beta ^*+ \alpha ^* \beta ) \right].
\end{equation}
Using again the solution (\ref{solu}) of identity (\ref{condw})
derived in the appendix, the function $f(\beta , \gamma )$ can
formally be written as
\begin{eqnarray}\!\!\!\!\!\!\!\!\!\!\!\!\!\!\!\!\!\!\!\!\!\!\!\!\!\!\!\!\!\!
f(\beta ,\gamma ) = \frac{4 \sigma ^4 }{\pi (1+ 2\sigma ^2)}
\,e^{\frac{1-2 \sigma ^2} {1+ 2\sigma ^2}|\beta |^2}\,e^{2\frac{1+2
\sigma ^2 -8 \sigma ^4}{(1+ 2\sigma ^2)^2 }|\gamma |^2} \int _{\mathbb
C}d^2 \xi \, e^{|\xi |^2 }\, e^{\xi (\beta - \frac {4 \sigma ^2 }{1+
2\sigma ^2}\gamma ) - \xi ^*(\beta ^* -\frac {4 \sigma ^2 }{1 +2\sigma
^2}\gamma ^*)}\;,\label{d2}
\end{eqnarray}
and thus the state (\ref{corcoh}) is an example of separable faithful
state.

\subsection{Example 3: product states}
Consider a product state 
\begin{equation}
R= \rho \otimes \sigma .
\end{equation}
The Wigner function is given by the product of the independent Wigner
functions for $\rho $ and $\sigma $
\begin{equation}
W_R(\alpha ,\beta )= W_\rho (\alpha ,\alpha ^*) W_\sigma (\beta ,\beta
^*) .
\end{equation}
Of course the state $R$ is not faithful, and in fact the condition
(\ref{condw}) can never be satisfied.
\subsection{Example 4: classical correlation between orthogonal states}
Consider the state
\begin{equation}
R=(1- \lambda )\sum _{n=0}^\infty \lambda ^n |n \rangle \langle n |
^{\otimes 2} \;,
\end{equation}
where $|n \rangle $ denotes the Fock state. From the relation
\cite{gla}
\begin{equation}
\langle n| D(\alpha )| n \rangle =e^{-\frac{|\alpha |^2}{2}}\,
L_n(|\alpha |^2) ,
\end{equation}
and the identity \cite{grad}
\begin{equation}
\sum_{n=0}^\infty \lambda ^n L_n (x)L_n(y)=\frac {1}{1-\lambda
}\,e^{-\frac{\lambda }{1-\lambda }(x+y) }\,I_0\left (2\frac{\sqrt {xy \lambda
}}{1-\lambda }\right )
 ,
\end{equation}
where $L_n(x)$ and $I_0(x)$ denote the $n$-th order Laguerre
polynomials and the $0$-order modified Bessel function, one obtains
the Wigner function
\begin{equation}
W_R(\alpha ,\beta )=\frac {4}{\pi^2}\, e^{-2 \frac{1+\lambda
    }{1-\lambda }(|\alpha |^2+|\beta |^2)}\, I_0 \left(\frac {8 \sqrt
    {\lambda }}{1-\lambda } |\alpha \beta |\right) .
\end{equation}
Condition (\ref{condw}) can never be satisfied, since there is no
dependence of the Wigner function on the phase of $\beta $.  In fact,
$\check R \equiv R$ is clearly not invertible, whence the state $R$ is
not faithful.
\section{Simplification for Gaussian states}
Unfortunately, it is often difficult to inspect condition
(\ref{condw}), since it holds more generally in a distribution
sense. For Gaussian bipartite states, however, it is possible to
derive a more practical condition in terms of the correlation matrix.
\par According to the derivation in the appendix, the term of the
Wigner function of a Gaussian bipartite state that is relevant for the
condition (\ref{condw}) is the factor of the form
\begin{equation}
g(\alpha ,\beta )=\exp[(A \alpha \beta  + B \alpha \beta ^*)+
  \hbox{h.c.}] \;,
\label{cons}
\end{equation}
In fact, as shown in the appendix, the condition (\ref{condw}) can be
satisfied iff
\begin{equation}
|A|^2-|B|^2 \neq 0 .\label{ab}
\end{equation}
In order to clarify the meaning of condition (\ref{ab}), it is useful
to consider the state $R$ in terms of the characteristic function
$\Gamma (\alpha ,\beta )= \hbox{Tr}[R D(\alpha )\otimes D(\beta )]$,
that corresponds to the Fourier transform of the Wigner function, and
hence will be Gaussian as well.  One has
\begin{equation}
R=\int _{\mathbb C}\frac {d^2 \alpha }{\pi}
\int _{\mathbb C}\frac {d^2 \beta }{\pi} \,\Gamma (\alpha ,\beta )
D^\dag (\alpha )\otimes D^\dag (\beta)
 .
\end{equation}
The operator $\check R$ then can be written as
\begin{equation}
\check{R} = \int _{\mathbb C}\frac {d^2 \alpha }{\pi}
\int _{\mathbb C}\frac {d^2 \beta }{\pi} \,\Gamma (\alpha ,\beta )
\dket{D^\dag (\beta )}\dbra {D^\dag (\alpha ^*)} .
\end{equation}
Similarly to Eq. (\ref{condw}), $\check R$ is invertible iff one can
find a function $f(\beta , \gamma  )$ such that
\begin{equation}
\int _{\mathbb C}d^2 \beta \,\Gamma (\alpha ,\beta ) \,f(\beta ,\gamma )= 
\delta ^{(2)}(\alpha -\gamma ) ,\label{condg} 
\end{equation}
and $\check R ^{-1}$ can be written as 
\begin{equation}
\check R ^{-1}= 
\int _{\mathbb C} 
d^2 \delta  \int _{\mathbb C} d^2 \gamma  \,f(\delta ,\gamma )
\dket{D^\dag (\gamma  ^*) }\dbra{D^\dag ( \delta  )} 
 .
\end{equation}
The same consideration before Eq. (\ref{cons}) applies here. The
condition (\ref{condg}) can be satisfied iff $|A|^2 - |B|^2 \neq 0$,
where $A$ and $B$ are the coefficients in front of the variables
$\alpha \beta $ and $\alpha \beta ^*$ of the characteristic
function. Using the identities
\begin{eqnarray}
&&A=\partial ^2 _{\alpha \beta }\Gamma (\alpha,\beta)  |_{\alpha = \beta
  =0} - \partial  _{\alpha }\Gamma (\alpha,\beta)  |_{\alpha = \beta
  =0} \, \partial  _{\beta }\Gamma (\alpha,\beta)  |_{\alpha = \beta =0}
\nonumber \\
&&A^*=\partial ^2 _{\alpha ^*\beta ^*}\Gamma (\alpha,\beta)  |_{\alpha = \beta
  =0} - \partial  _{\alpha ^*}\Gamma (\alpha,\beta)  |_{\alpha = \beta
  =0} \, \partial  _{\beta ^*}\Gamma (\alpha,\beta)  |_{\alpha =
  \beta =0}
\nonumber \\ 
&&B=\partial ^2 _{\alpha \beta ^*}\Gamma (\alpha,\beta)  |_{\alpha = \beta
  =0} - \partial  _{\alpha }\Gamma (\alpha,\beta)  |_{\alpha = \beta
  =0} \, \partial  _{\beta ^*}\Gamma (\alpha,\beta)  |_{\alpha = \beta =0}
\nonumber \\ 
&&B^*=\partial ^2 _{\alpha ^*\beta }\Gamma (\alpha,\beta)  |_{\alpha = \beta
  =0} - \partial  _{\alpha ^*}\Gamma (\alpha,\beta)  |_{\alpha = \beta
  =0} \, \partial  _{\beta }\Gamma (\alpha,\beta)  |_{\alpha = \beta =0},
\end{eqnarray}
the condition of faithfulness can be restated in terms of the
correlation matrix as follows:
\paragraph{\bf Necessary and sufficient condition for faithfulness of Gaussian
states:} a bipartite Gaussian state is faithful iff the following
condition on the correlations is satisfied
\begin{equation}
\chi(R)\doteq\langle  \Delta a^\dag b^\dag \rangle  
\langle  \Delta ab \rangle   + \langle  \Delta a^\dag b \rangle  
\langle  \Delta ab ^\dag  \rangle \neq 0,
\end{equation}
where for any two operators $P$ and $Q$ 
\begin{equation}
\langle \Delta PQ\rangle \doteq\langle PQ\rangle-\langle
P\rangle\langle Q\rangle \,.
\end{equation}

\par In terms of the quadratures $X_c =(c + c^\dag )/2$ and
$Y_c=(c-c^\dag )/(2i)$ of the modes $c=a,b$, the correlation $\chi(R)$
can be rewritten as
\begin{equation}
\chi(R)= \frac 12 \left(\langle \Delta X_a X _b\rangle ^2 + \langle
\Delta Y_a Y_b \rangle ^2 +\langle \Delta X_a Y _b \rangle ^2 +\langle
\Delta Y_a X_b \rangle ^2 \right).\label{corrXY}
\end{equation}
In the examples 1---4 given in the previous section, one has
  $\chi(R)=\frac{\lambda ^2}{(1-\lambda ^2)^2}\,,\sigma ^4\,, 0\,$,
  and $0$, respectively (notice, however, that the state in example 4
  is not Gaussian).  Using Eq. (\ref{corrXY}) the condition $\chi(R)
  \neq 0$ shows that bipartite Gaussian states are always faithful,
  except when they are product states ($\chi(R)$ is the sum of
  nonnegative terms which all vanish when there is no correlation
  between the two modes).  A rigorous relation bewteen statistical
  errors that affect tomographic reconstructions and the strength of
  correlations is in order, but beyond the scope of this paper.

\section{Conclusions}
Tomographically faithful states are a necessary ingredient for
tomography of quantum operations and for complete quantum calibration
of measuring apparatuses. In this paper we have provided a complete
classification of two-mode faithful states in terms of the Wigner
function of the state. This classification has been derived from the
general faithfulness condition resorting to the invertibility of a
special operator associated to the state. Some examples of faithful
states have been presented, both entangled and separable, along with
examples of not faithful states.  For two-mode Gaussian states we have
shown that faithfulness is simply equivalent to nonvanishing
correlations between the modes.  

\par We conclude by noticing that the actual statistical efficiency of
a faithful state in the tomography of a quantum operation in infinite
dimensions is connected to the increase of the singular values of the
unbounded operator $\check R^{-1}$. Such unboundedness is responsible
of the increasingly large statistical errors in the Fock
representation of the quantum operation, accounting for the finite
experimental data sample used to infer information on a infinite set
of matrix elements of the quantum operation. As a rule of thumb, the
statistical efficiency increases for greater correlation $\chi(R)$.
\par Finally, it is worth mentioning that the framework of {\em
quantum images} \cite{qph} bears a strict analogy with that of the
quantum tomography of a channel using an input bipartite state, with
the role of the channel here played by the density contour of the
image analyzed by one of the twin beams from parametric downconversion
of vacuum. Clearly, when the state is faithful one has quantum imaging
on the other beam, and our result is consistent with the recent
demonstration \cite{qph} that entanglement is not necessary for
quantum imaging. In particular, the thermal state split by a beam
splitter in Eq. (\ref{corcoh}) is suitable for quantum imaging.  \ack
This work has been sponsored by INFM through the project
PRA-2002-CLON, and by EC and MIUR through the cosponsored ATESIT
project IST-2000-29681 and Cofinanziamento 2003.
\section*{Appendix} 
We show that for a function of the form 
\begin{equation}
g(\alpha ,\beta )= h(\alpha )\,k(\beta ) \,e^{
A\alpha\beta+ A^*\alpha^*\beta^*}\, 
e^{B\alpha\beta^* +B^* \alpha^*\beta}
\;,
 \end{equation}
with $|A|\neq |B|$, and both $h$ and $k$ (generally not
analitycal) invertible functions, 
the following function
\begin{eqnarray}
&&\!\!\!\!\!\!\!\!\!\!\!\!\!\!\!\!\!\!\!\!\!\!\!\!\!\!
f(\beta,\gamma)=
\frac{(|A|^2 -|B|^2)^2}{\pi ^3}
\,k^{(-1)}(\beta )\,h^{(-1)}(\gamma 
)\, e^{-(|A|^2+|B|^2)|\gamma |^2 -(AB \gamma ^2 + A^*B^*\gamma
  ^{*2})}\, e^{-|\beta |^2}\nonumber \\& & 
\!\!\!\!\!\!\!\!\!\!\!\!\!\!\!\!\!\!\!\!\!\!\!\!\!\!
\times 
\int d^2 \xi \,
e^{(|A|^2+|B|^2)|\xi |^2 + (A^*B^*\xi ^2 + AB\xi ^{*2})}\,
e^{(|A|^2-|B|^2)(\xi \gamma -\xi^* \gamma ^*)}\,
e^{\beta (A\xi^* +B^*\xi )- \beta ^*(A^*\xi + B\xi ^* )}\;,  
\label{solu}
\end{eqnarray}
satisfies the identity 
\begin{equation}
\int _{\mathbb C}d^2 \beta \,g(\alpha ,\beta ) \,f(\beta ,\gamma )= 
\delta ^{(2)}(\alpha -\gamma ) \;.
\label{condwap}
\end{equation}
The function $f(\beta ,\gamma )$ should be treated as a distribution,
in the sense that the integral in $\xi $ has to be performed after the
integration on $\beta $ of Eq. (\ref{condwap}).  \par One has
\begin{eqnarray}
&&\!\!\!\!\!\!\!\!\!\!\!\!\!\!\!\!\!\!\!\!\!\!\!\!\!\!\!\!\!\!
\int _{\mathbb C}d^2 \beta \,g(\alpha ,\beta ) \,f(\beta ,\gamma )= 
\frac{(|A|^2 -|B|^2)^2}{\pi ^3}\,h(\alpha )\,h^{(-1)}(\gamma )\,
e^{-(|A|^2+|B|^2)|\gamma |^2 - (AB\gamma ^2 + A^*B^*\gamma
  ^{*2})}
\nonumber \\& & 
\!\!\!\!\!\!\!\!\!\!\!\!\!\!\!\!\!\!\!\!\!\!\!\!\!\!\!\!\!\!
\times 
\int d^2 \xi \,
e^{(|A|^2+|B|^2)|\xi |^2 +(A^*B^*\xi ^2 + AB \xi ^{*2})}\,
e^{(|A|^2-|B|^2)(\xi \gamma -\xi^* \gamma ^*)}\nonumber \\& & 
\!\!\!\!\!\!\!\!\!\!\!\!\!\!\!\!\!\!\!\!\!\!\!\!\!\!\!\!\!\!
\times \int d^2 \beta  \, e^{-|\beta |^2}\,
e^{\beta (A\xi ^* +B^* \xi +A \alpha +B ^*\alpha ^* )- \beta ^*
(A^*\xi  +B \xi^*  -A ^* \alpha ^* - B \alpha  )}
\nonumber \\& & \!\!\!\!\!\!\!\!\!\!\!\!\!\!\!\!\!\!\!\!\!\!\!\!\!\!\!\!\!\!
=\frac{(|A|^2 -|B|^2)^2}{\pi ^2}\,h(\alpha )\,h^{(-1)}(\gamma )\,
e^{(|A|^2+|B|^2)(|\alpha |^2 -|\gamma |^2) + AB 
(\alpha ^2 -\gamma ^2 )+ A^*B^*
(\alpha ^{*2}- \gamma ^{*2})}\nonumber \\& &
\!\!\!\!\!\!\!\!\!\!\!\!\!\!\!\!\!\!\!\!\!\!\!\!\!\!\!\!\!\!
\times 
\int d^2 \xi \,
e^{(|A|^2-|B|^2)[\xi (\gamma -\alpha )- \xi^* (\gamma ^* - \alpha ^*)]}
\nonumber \\& &
\!\!\!\!\!\!\!\!\!\!\!\!\!\!\!\!\!\!\!\!\!\!\!\!\!\!\!\!\!\!
=
\delta ^{(2)}(\alpha -\gamma )
\;,\end{eqnarray}
where the integral in $d^2 \beta $ has been performed by using the
identity
\begin{eqnarray}
\int _{\mathbb C}d^2 \beta \,e^{-\frac {|\beta |^2}{\sigma ^2}}\,
e^{\beta \alpha ^* - \beta ^* \gamma }= \pi \sigma ^2 \,e^{-\sigma ^2
  \alpha ^* \gamma }\;.\label{}
\end{eqnarray}


\section*{References}
\begin{thebibliography}{99}
\bibitem{tomo_channel} G. M. D'Ariano and P. Lo Presti,
Phys. Rev. Lett. {\bf 86}, 4195 (2001)
\bibitem{calib} G. M. D'Ariano, P. Lo Presti,  
and L. Maccone, Phys. Rev. Lett. {\bf 93}, 250407 (2004).  
\bibitem{prl} G. M. D'Ariano and P. Lo Presti, 
Phys. Rev. Lett. {\bf 91}, 047902 (2003).
\bibitem{phs} {\em The Physics of Phase Space}, edited by Y. S. Kim
and W. W. Zachary (Springer, Berlin, 1986).
\bibitem{gard} C. W. Gardiner, {\em Quantum Noise} (Springer, Berlin,
1991).
\bibitem{Kraus}K. Kraus, {\em States, Effects and Operations}, Lecture
Notes in Physics Vol. 190 (Springer, Berlin, 1983).
\bibitem{pla} G. M. D'Ariano, P. Lo Presti, and M. F. Sacchi,
Phys. Lett. {\bf 272}, 32 (2000).
\bibitem{gla}K. E. Cahill and R. J. Glauber, Phys. Rev. {\bf 177},
1857 (1969).
\bibitem{grad} I. S. Gradstein and I. M. Ryzhik, {\em Table of
Integrals, Series, and Products} (Academic Press, New York, 1980).
\bibitem{qph} D. Magatti, F. Ferri, A. Gatti, M. Bache, E. Brambilla,
and L. A. Lugiato, Phys. Rev. Lett. {\bf 94}, 183602 (2005).
\end{thebibliography}
\end{document}